\documentclass[5p,authoryear,x11names]{elsarticle}
\usepackage{graphicx}
\usepackage{amssymb}
\usepackage{mathabx}
\usepackage{float}
\usepackage{hyperref}
\usepackage{tikz}
\usepackage{listings}
\usepackage{algpseudocode}
\usepackage{algorithm}
\usepackage{lscape}
\usepackage{longtable,booktabs,tabu}

\newcommand{\cross}{\ding{54}}
\newcommand{\tick}{\ding{52}}
 
\title{\texttt{HMFcalc} : An Online Tool for Calculating Dark Matter Halo Mass 
Functions}
\author{S.~G.~Murray$^{1,2}$,
         C.~Power$^{1,2}$ \& A.~S.~G.~Robotham$^{1}$\\
        $^1$ ICRAR, University of Western Australia, 35 Stirling Highway,
        Crawley, Western Australia 6009, Australia\\
        $^2$ ARC Centre of Excellence for All-Sky Astrophysics (CAASTRO)}

\begin{document}

\begin{abstract}

The dark matter halo mass function (HMF) is a characteristic property of 
cosmological structure formation models, quantifying the number density 
of dark matter haloes per unit mass in the Universe. A key goal of 
current and planned large galaxy surveys is to measure the HMF and to 
use it to test theories of dark matter and dark energy. We present a new 
web application for calculating the HMF -- the frontend \verb|HMFcalc| 
and the engine \verb|hmf|. \verb|HMFcalc| has been designed to be 
flexible, efficient and easy to use, providing observational and 
theoretical astronomers alike with the means to explore standard 
functional forms of the HMF or to tailor their own. We outline the 
theoretical background needed to compute the HMF, we show how it has 
been implemented in \verb|hmf|, and finally we provide worked examples 
that illustrate \verb|HMFcalc|'s versatility as an analysis tool.

\end{abstract}

\maketitle

%% main text
\section{Introduction}
\label{sec:intro}

There is now a wealth of compelling observational evidence that we live in
a Universe whose matter content is predominantly dark 
\citep[$\sim$84\%; cf.][]{PlanckCollaboration2013} and non-baryonic in 
nature \citep[cf.][]{Bergstrom2000}. Our theories of cosmological structure 
formation predict that this dark matter
clusters into massive gravitationally bound structures called haloes. 
The dark matter halo mass function (hereafter HMF) quantifies the number of these 
haloes per unit comoving volume of the Universe as a function of their 
mass. The HMF is sensitive to the cosmological parameters, primarily the 
mass-energy density of dark matter $\Omega_c$ and dark energy $\Omega_{\Lambda}$ 
\citep[e.g.][]{Murray2013}, but it also depends on the nature of the dark matter. The 
standard Cold Dark Matter (CDM) model predicts an HMF in which the number 
of haloes increases with decreasing halo mass $M$ approximately as 
$M^{-1.8}$ \citep[e.g.][]{Lukic2007,Bhattacharya2011}, whereas viable Warm Dark Matter 
(WDM) models predict fewer haloes than the CDM model at low masses\citep[e.g.][]{Schneider2013,Pacucci2013}. The potential 
of the HMF as a probe of dark matter and dark energy is widely recognised 
\citep[e.g.][]{Tinker2008,Vikhlinin2009} and 
is one of the key science drivers of current and planned future galaxy 
surveys \citep[][]{Driver2011,Pierre2011}.

Cosmological $N$-body simulations are now established as the tool for 
studying the HMF \citep[cf. the recent review by][]{Knebe2013}, but the 
information contained in a simulation is usually
distilled and recast in a more compact form. Usually this is the comoving 
number density of haloes per unit logarithm of the halo mass $M$, 
\begin{equation}
\label{eq:hmf}
  \frac{dn}{d\ln M} = M\cdot \frac{\rho_0}{M^2} f(\sigma) \left|\frac{d\ln\sigma}{d\ln M}\right|;
\end{equation}
here $\sigma$ and $\rho_0$ are the cosmology-dependent mass variance and mean 
density and $f(\sigma)$ represents the functional form that defines a particular
HMF fit. Eq~\ref{eq:hmf} is not difficult to compute, but neither is it 
straightforward. We have developed an easy-to-use and flexible 
web-application to compute Eq~\ref{eq:hmf}, which we call
\verb|hmf|, and its frontend \verb|HMFcalc| which can be accessed
at \url{http://hmf.icrar.org}. \verb|HMFcalc| can be used in
a number of ways, including as
\begin{itemize}
  
\item a standard against which to check one's own code;
  
\item an easy-to-use interface to generate HMFs against 
  which to check observational/simulations data; and
  
\item a visually intuitive way to explore the effects of cosmology 
  on the HMF.
  
\end{itemize} 
The objective of this paper is to present a detailed overview of 
\verb|hmf| and \verb|HMFcalc|, describing its implementation and the
underlying philosophy for this approach, as well as providing some
worked examples that illustrate its usefulness and versatility.

The paper is structured as follows. In \S\ref{sec:theory} we provide the
theoretical background necessary to compute the HMF, setting out a
compilation of HMF fitting functions drawn from the literature and 
demonstrating how the HMF differs in CDM and WDM
models. In \S\ref{sec:implementation} we describe our implementation of 
\verb|hmf| and \verb|HMFcalc| and discuss the algorithms and methods used.
In \S\ref{sec:application} we present some worked examples using \verb|HMFcalc|
and finally in \S\ref{sec:future} we summarise our plans for extending
\verb|HMFcalc|'s functionality and usability.

\section{The Halo Mass Function (HMF)}
\label{sec:theory}
The HMF quantifies the number of dark matter haloes per unit
mass per unit comoving volume of the Universe,
\[
\frac{dn}{d\ln M} = M\cdot \frac{\rho_0}{M^2} f(\sigma) \left|\frac{d\ln\sigma}{d\ln M}\right|
\]
where $f(\sigma)$ is the fitting function that we shall return to 
shortly, $\rho_0$ is the mean density of the Universe and $\sigma$ is 
the rms variance of mass within a sphere of radius $R$ that contains
mass $M$,
\begin{equation}
  \label{eq:MandR}
  M = \frac{4\pi\rho_0}{3}R^3.
\end{equation}
The mass variance is calculated via the integral,
\begin{equation}
	\label{eq:massvariance}
	\sigma^2(R) = \frac{1}{2\pi^2}\int_0^\infty{k^2P(k)W^2(kR)dk}
\end{equation}
where $k=2\pi/\lambda$ is the wavenumber corresponding to the wavelength $\lambda$ of the 
fluctuation, $P(k)$ is the linear power spectrum (see below) and $W(kR)$ is 
the Fourier transform of the top-hat window function\footnote{Note that 
  there is freedom in the choice of window function; 
  we follow the convention in which the window function 
  is a top-hat in real space, but we could adopt a Gaussian in real space, 
  or a top-hat in Fourier space. The 
  dependence of $\sigma$ on the choice of window function is well known and the 
  implications have been explored in studies such as \citet{Schneider2013}.
  We shall return to this point in \S\ref{sec:future}.}
(a sphere with sharp edges in real space), i.e.
\begin{equation}
	\label{eq:window}
 	W(kR) = \frac{3\left[\sin(kR) - kR\cos(kR)\right]}{(kR)^3}.
\end{equation}
The right-most factor of Eq \ref{eq:hmf} can be written as
\begin{equation}
  \label{eq:dlndlnm}
  \frac{d\ln\sigma}{d\ln M} = \frac{3}{2\sigma^2\pi^2R^4}\int_0^\infty \frac{dW^2(kR)}{dM}\frac{P(k)}{k^2}dk
\end{equation}
using Eq~\ref{eq:MandR} to relate $M$ to $R$, and writing
\begin{eqnarray}
  \label{eq:windowderiv}
  \frac{dW^2}{dM}&=&\left[\sin(kR) - kR\cos(kR) \right] \times \nonumber \\
                  & &\left[\sin(kR)\left(1-\frac{3}{(kR)^2}\right)+3\frac{\cos(kR)}{kR}\right].
\end{eqnarray}

\begin{figure}
  \centering
  \includegraphics[width=\linewidth]{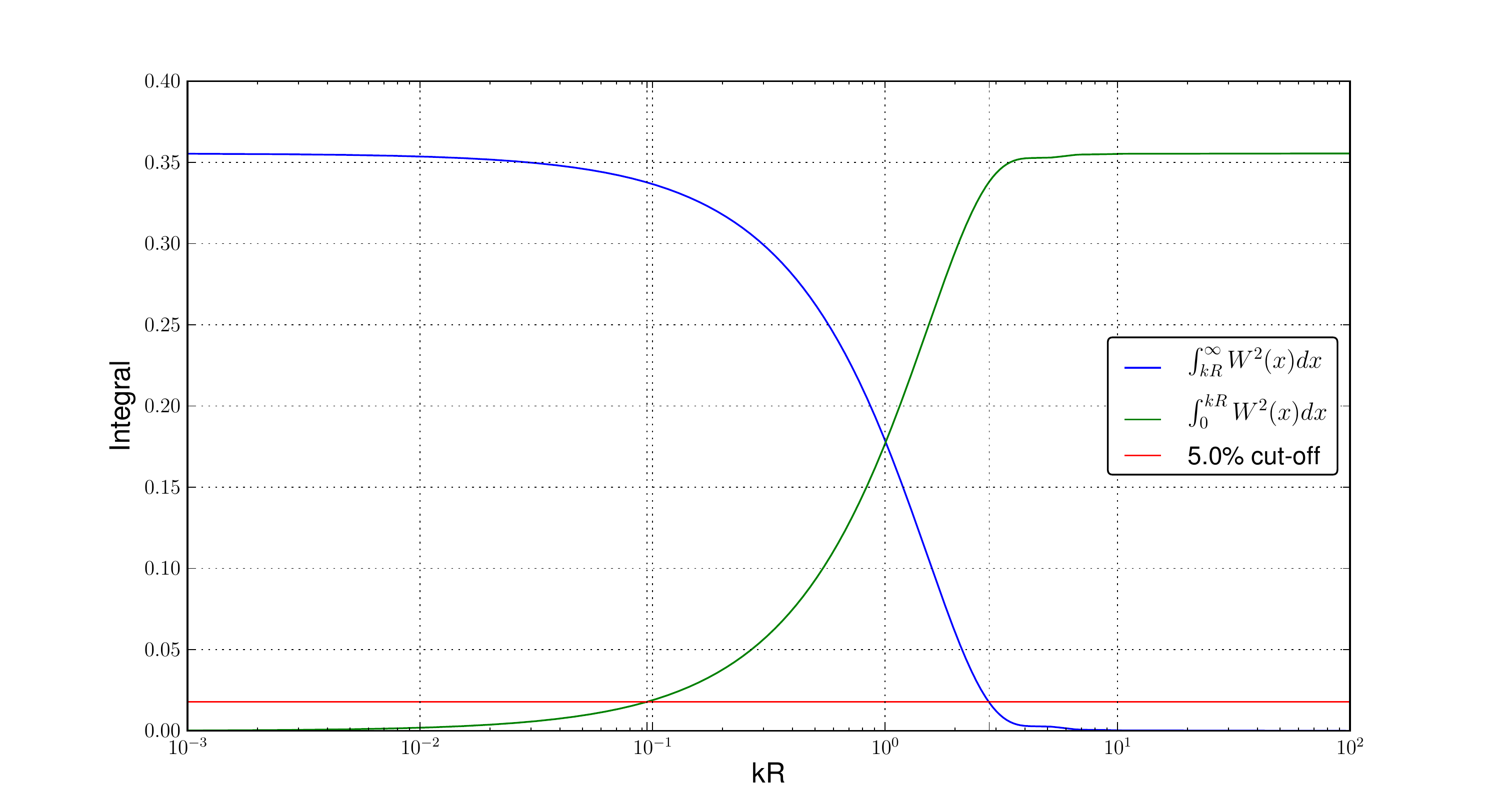}
  \caption{The integral of the window function gives an upper limit of 0.1 on the minimum 
    of the $kR$ range, and a lower limit of 3 on the maximum of the $kR$ range, with a 95\% 
    contribution of the total integral. }
  \label{fig:windows}
\end{figure}

The window function and its derivative are functions of the 
product $kR$, but we evaluate Eqs \ref{eq:massvariance} and 
\ref{eq:dlndlnm} by integrating over $k$. For this reason care must be 
taken when solving the integrals numerically to ensure that the results 
are converged. We demonstrate why in Fig~\ref{fig:windows}, where we 
plot $\int_0^{kR} W^2(x)dx$ and $\int_{kR}^\infty W^2(x)dx$. The
integral $\int_0^{kR} W^2(x)dx$ allows us to identify an upper limit on the
minimum $kR$ required for convergence; we want the range of $kR$ 
for any $R$ to have a minimum that bounds the non-zero parts of the 
function. As a rule of thumb, we identify this as the value of $kR$ for 
which 95\% of $\int_0^{\infty} W^2(x)dx$ is kept; from Fig~\ref{fig:windows},
this corresponds to $kR = 0.1$ and so we have the condition,
\begin{equation}
	\label{eq:maxkrrule}
	k_{\mathrm{min}}R_{\mathrm{max}} < 0.1
\end{equation}
Repeating this exercise for $\int_{kR}^\infty W^2(x)dx$, we deduce
\begin{equation}
	\label{eq:minkrrule}
	k_{\mathrm{max}}R_{\mathrm{min}} > 3.
\end{equation}
These bounds on $kR$ are implemented in \verb|hmf| by default, which issues 
a warning if they are not obeyed. Usually the limits on $k$ to which the 
power spectrum is extrapolated far exceed these rules for any sensible mass 
range, but instances can arise where more stringent limits on $k$ are required 
(cf. \S\ref{sec:finitesize}).

The linear power spectrum $P(k)$ characterises the distribution of matter 
density perturbations as a function of wavenumber $k$; formally
it is computed by averaging over $\left|\hat{\delta(\vec{k})}\right|$, the square of the
Fourier-transformed real-space density fluctuations 
$\delta(\vec{x})=\rho(\vec{x})/\bar{\rho}-1$, where $\vec{x}$ and $\vec{k}$ are 
the spatial coordinate and wave vector respectively. We express it as
\begin{equation}
	\label{eq:powerfromtransfer}
	P(k) = Ak^n T^2(k),
\end{equation}
where $T(k)$ is the transfer function, $A$ is the normalisation constant and
$n$ is the spectral index. We follow convention and use the cosmological 
parameter $\sigma_8$, which measures the mass variance on a scale of $8 h^{-1}$Mpc, 
to calculate $A$. The primordial power spectrum, imprinted during the epoch of
inflation during the first moments after the Big Bang, is expected to have a form
$P(k) \propto k^n$. The transfer function quantifies how this primordial form is 
modified on different scales, and it is particularly sensitive to the nature of
the dark matter and the baryon density parameter $\Omega_{\rm b}$. We use the 
public \textit{Code for Anisotropies in the Microwave 
Background} (CAMB) \citep{Lewis2000} to compute our transfer functions.

\subsection{Fitting Functions}
\label{sec:refinedfittingfunctions}
As noted in the introduction, it is the fitting function $f(\sigma)$ that 
defines a particular HMF fit. Early analytical work by \citet{Press1974} 
(hereafter PS) and \citet{Bond1991} established a simple form for $f(\sigma)$ 
by assuming that haloes form by spherical collapse, finding
\begin{equation}
  f(\sigma) = \sqrt{\frac{2}{\pi}}\frac{\delta_c}{\sigma}\exp\left(-\frac{\delta_c^2}{2\sigma^2}\right),
\end{equation}
where $\delta_c \simeq 1.686$ is the critical overdensity for spherical collapse.

However, $N$-body simulations of cosmological structure formation 
have revealed that the PS form underestimates the abundance of
higher mass haloes and overestimates the abundance of lower mass haloes.
\citep[e.g.][]{Sheth2001,White2002,Lukic2007}. \citet{Sheth2001}
(hereafter ST) explored an extension to the PS formalism by considering 
ellipsoidal rather than spherical collapse and obtained a form for the mass
function that is identical to Eq~\ref{eq:hmf} but with a modified $f(\sigma)$.
Subsequent studies have largely adopted the same philosophical approach of 
assuming that the HMF can be expressed in the form of Eq~\ref{eq:hmf}
and using $f(\sigma)$ to characterise the HMF. 

\begin{figure}
  
  \centering
  \includegraphics[width=\linewidth, trim=2cm 0cm 3.0cm 0cm, clip=True]{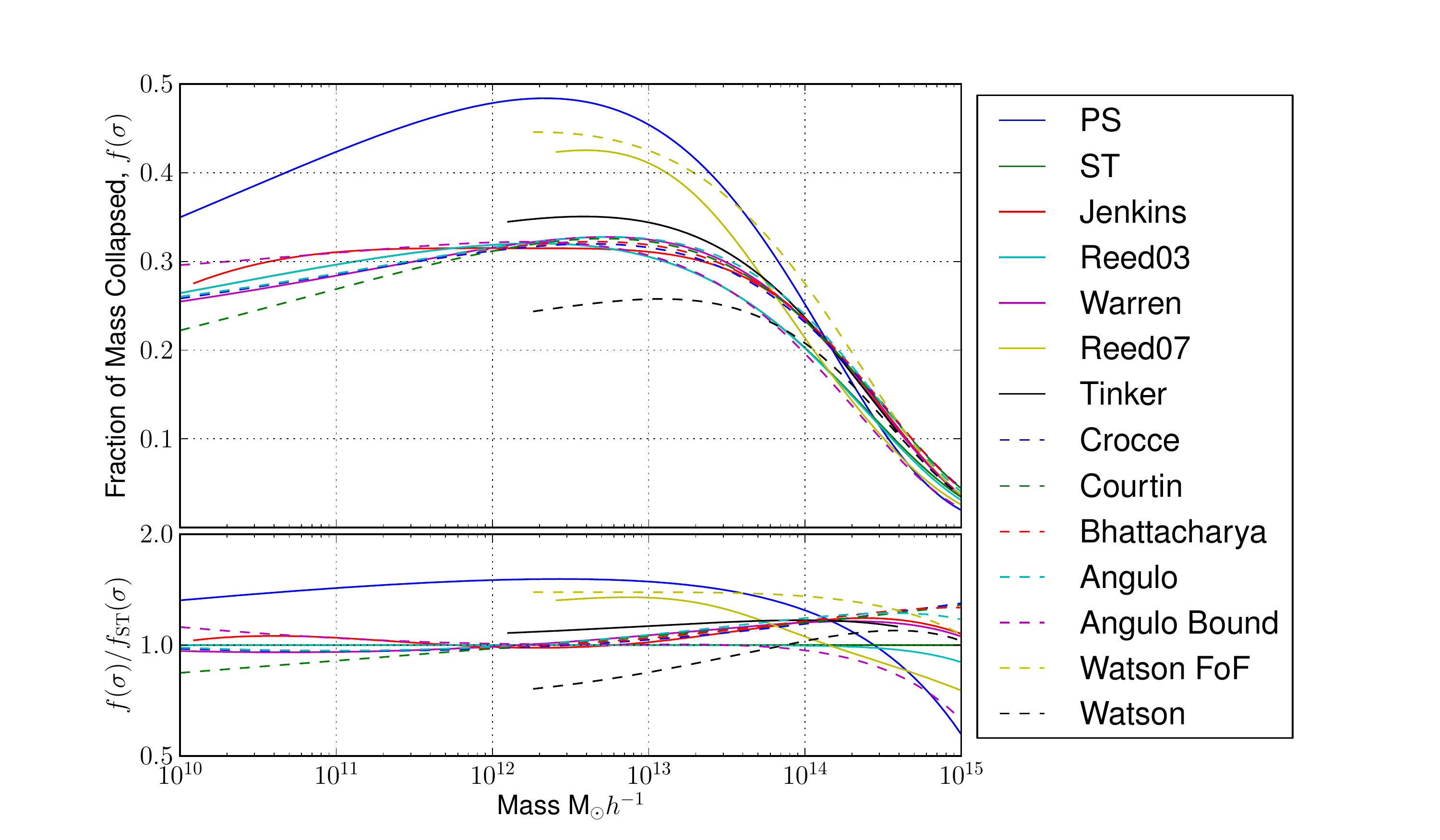}
  \caption{Upper: all fitting functions at redshift zero over a large mass range (limits placed as appropriate on each function). Lower: each fitting function divided by the Sheth-Tormen fit.}
  \label{fig:fsigma}
\end{figure}

Table \ref{tab:fittingfunctions} provides a concise summary of the forms for 
$f(\sigma)$ that have appeared in the literature to date and which are included 
in \verb|HMFcalc|, and we list also the cosmology and mass and redshift ranges over 
which the fits have been made\footnote{Note the parameters $A,a,b,c$ for the Tinker fit are actually dependent on $\Delta_{\rm vir}$. See \cite{Tinker2008} or the code for details.}. Each of these fitting functions are plotted in 
Fig \ref{fig:fsigma} for comparison.

\begin{table*}
\centering
  \caption[]{Compilation of Fitting Functions}
  \label{tab:fittingfunctions}	
  \resizebox{!}{11.4cm}{
    \begin{tabu} to \linewidth{X[-2.25l]X[9.0l]X[3.4c]X[1.5c]X[3l]} 
      
      \toprule[0.05cm]
      \textsc{Ref.} & \textsc{Fitting Function $f(\sigma)$} &\textsc{Mass Range} & \textsc{Redshift Range} & \textsc{Cosmology Fitted}\\ \midrule[0.04cm] 
      
      \citet{Press1974} & $f_{\rm PS}(\sigma) = \sqrt{\frac{2}{\pi}}\frac{\delta_c}{\sigma}\exp\left[-\frac{\delta_c^2}{2\sigma^2}\right]$ &  -- &-- & --\\ \midrule
      
      \citet{Sheth2001} & $f_{\rm ST}(\sigma) =A\sqrt{\frac{2a}{\pi}}\left[1+\left(\frac{\sigma^2}{a\delta_c^2}\right)^p\right]\frac{\delta_c}{\sigma}\exp\left[-\frac{a\delta_c^2}{2\sigma^2}\right]$, \newline $A=0.3222$, $a=0.707$, $p=0.3$. &  -- & -- & Einstein-de Sitter\\ \midrule
      
      \citet{Jenkins2001} & $f_{\rm J}(\sigma) = 0.315 \exp \left[|\ln \sigma^{-1}+0.61|^{3.8}\right]$ &  $-1.2 < \ln \sigma^{-1} < 1.05$ & 0 -- 5 & $\tau$CDM, $\Lambda$CDM\\ \midrule
      
      \citet{Reed2003} & $f_{\rm R03}(\sigma) =f_{ST}(\sigma) \exp\left[\frac{-0.7}{\sigma \cosh(2\sigma)^5}\right]$ &  $-1.7< \ln \sigma^{-1} <0.9$& 0 -- 15 & $\Omega_M=0.3$, $\Omega_\Lambda = 0.7$\\ \midrule
      
      \citet{Warren2006} & $f_{\rm W}(\sigma) =0.7234 \left(\sigma^{-1.625}+0.2538\right) \exp\left[\frac{-1.1982}{\sigma^2}\right]$ &  $10^{10}{\rm M}_{\Sun} < M < 10^{15}{\rm M}_{\Sun}$ & 0 &  $\Lambda$CDM: WMAP1\\ \midrule		
      
      \citet{Reed2007} & 
      $f_{\rm R07}(\sigma)=\nu \exp\left[-\frac{ca\frac{\delta_c^2}{\sigma^2}}{2} - \frac{0.03\left(\frac{\delta_c}{\sigma}\right)^{0.6}}{(n_{\rm eff}+3)^2}\right] $
      $\times A\sqrt{\frac{2a}{\pi}} \left[1+\left(\frac{1}{\frac{\delta_c^2}{\sigma^2} a}\right)^p + 0.6G_1(\sigma)+0.4G_2(\sigma)\right]$ \newline 
      $n_{\mathrm{eff}} = 6\frac{d \log \sigma^{-1}}{d \log M} -3$, 
      $G_1(\sigma) = \exp \left[-\frac{\ln(\sigma^{-1}-0.4)^2}{0.72}\right]$,
      $G_2(\sigma) = \exp \left[-\frac{\ln(\sigma^{-1}-0.75)^2}{0.08}\right]$  
      &  $-1.7 < \ln \sigma^{-1} < 0.9$ & 0 -- 30 & $\Lambda$CDM: WMAP1\\ \midrule

      \citet{Tinker2008} & 
      $f_{\rm T}(\sigma,z) =A\left(\left(\frac{b}{\sigma}\right)^a + 1\right) \exp\left[-\frac{c}{\sigma^2}\right]$, \newline
      $A = 0.186\left(1+z\right)^{-0.14}$,	$a = 1.47\left(1+z\right)^{-0.06}$, $	b = 2.57\left(1+z\right)^{-\alpha}$,$c = 1.19$,\newline
      $\alpha = \exp\left[-\left(\frac{0.75}{\ln(\Delta_{\mathrm{vir}}/75)}\right)^{1.2}\right]$& 
      $-0.6 < \ln \sigma^{-1} < 0.4$ & 0 -- 2.5 & $\Lambda$CDM: WMAP1, WMAP3+\\ \midrule		
      
      \citet{Crocce2010} & 
      $f_{\rm Cr}(\sigma) =A\left(\sigma^{-a} + b\right) \exp\left[-\frac{c}{\sigma^2}\right]$, \newline
      $A = 0.58\left(1+z\right)^{-0.13}$,	$a = 1.37\left(1+z\right)^{-0.15}$, $	b = 0.3\left(1+z\right)^{-0.084}$, $c = 1.036\left(1+z\right)^{-0.024}$& 
      $10^{10.5}{\rm M}_{\Sun} < M < 10^{15.5}{\rm M}_{\Sun}$ & 0 -- 2 &$(\Omega_M,\Omega_\Lambda,n,h,\sigma_8)$ $=(0.25,0.75,$ $0.95,0.7,0.8)$\\ \midrule			
      
      \citet{Courtin2010} & 
      $f_{\rm Co}(\sigma) =f_{\rm ST}(\sigma)$, \newline
      $A = 0.348$,	$a = 0.695$, $p = 0.1$& 
      $-0.8 < \ln \sigma^{-1} < 0.7$& 0  &$\Lambda$CDM: WMAP5\\ \midrule
      
      \citet{Bhattacharya2011} & 
      $f_{\rm B}(\sigma,z) = A\sqrt{\frac{2}{\pi}}\exp\left[-\frac{a\delta_c^2}{2\sigma^2}\right]\left[1+\left(a\frac{\delta_c^2}{\sigma^2}\right)^{-p}\right]\left(\frac{\delta_c^2}{\sigma^2}\sqrt{a}\right)^q$, \newline
      $A = 0.333\left(1+z\right)^{-0.11}$,	$a = 0.788\left(1+z\right)^{-0.01}$, $	p = 0.807$, $q =1.795$& 
      $10^{11.8}{\rm M}_{\Sun} < M < 10^{15.5}{\rm M}_{\Sun}$& 0 -- 2 & $w$CDM+ \\ \midrule	
      
      \citet{Angulo2012} & 
      $f_{\rm A}(\sigma) =A\left[\left(\frac{b}{\sigma}\right)^a+1\right] \exp\left[-\frac{c}{\sigma^2}\right]$, \newline
      $\left(A,a,b,c\right) = \left(0.201,1.7,2.08,1.172\right)$ or $\left(A,a,b,c\right)_{\rm SUB} = \left(0.265,1.9,1.675,1.4\right)$& 
      $10^{8}{\rm M}_{\Sun} < M < 10^{16}{\rm M}_{\Sun}$& 0 & $\Lambda$CDM: WMAP1 \\ \midrule	
      
      \citet{Watson2012} & 
      $f_{\rm W_{FOF}}(\sigma,z) = f_{\rm T}(\sigma,z)$, \newline
      $A = 0.282$,	$a = 1.406$, $	b = 2.163$, $c = 1.21$& 
      $-0.55 < \ln \sigma^{-1} < 1.31$ & 0 -- 30 & $\Lambda$CDM: WMAP5\\ \midrule	
      
      \citet{Watson2012} & 
      $f_{\rm W_{SO}}(\sigma,z) = \Gamma(\Delta, \sigma, z)f_{\rm T}(\sigma,z)$, \newline
      $\left(A,a,b,c\right)_{z=0} = \left(0.194,2.267,1.805,1.287\right)$, 
      $\left(A,a,b,c\right)_{z>6} = \left(0.563,874,3.810,1.453\right)$,
      $\left(A,a,b,c\right)_{0<z<6} = \Omega_M(z)\times(1.907(1+z)^{-3.216}+0.074$,
      $3.136(1+z)^{-3.058}+2.349$, $5.907\times(1+z)^{-3.599}+2.344$, $1.318)$, \newline
      $\Gamma(\Delta,\sigma,z) = C(\Delta)\left(\frac{\Delta}{178}\right)^{d(z)}\exp \left[\frac{p\left(1-\frac{\Delta}{178}\right)}{\sigma^q}\right]$, \newline
      $C(\Delta) = 0.947 \exp\left[0.023\left(\frac{\Delta}{178}-1\right)\right]$,
      $d(z) = -0.456\Omega_M(z)-0.139$,
      $p = 0.072$,
      $q = 2.130$.& 
      $-0.55 < \ln \sigma^{-1} < 1.05$ ($z=0$), \newline $-0.06 < \ln \sigma^{-1} < 1.024$ ($z>0$)  & 0 -- 30 & $\Lambda$CDM: WMAP5\\ %\midrule[0.04cm]
      \bottomrule[0.05cm]
    \end{tabu}
  }
\end{table*}

\subsection{Redshift Dependence}
\label{sec:evolution}

In general it is assumed that the form of the fitting function $f(\sigma)$ is 
insensitive to redshift, and so it is through the rms mass variance $\sigma(M,z)$ 
that redshift dependence is captured. The evolution of $\sigma(M,z)$ is governed 
by the growth factor,
\begin{equation}
  \label{eq:dofz}
  d(z) = \frac{D^+(z)}{D^+(z=0)}.
\end{equation}
where $D^+(a)$ is 
\begin{equation}
  \label{eq:Dplus}	
  D^+(z) = \frac{5\Omega_m}{2}\frac{H(z)}{H_0}\int_z^{\infty}{\frac{(1+z')dz'}{[H(z')/H_0]^3}},
\end{equation}
and $H_0$ is the present-day Hubble parameter with
\begin{equation}
  \label{eq:HofA}
  H(z) = H_0\sqrt{\Omega_m (1+z)^3 + (1-\Omega_m)}
\end{equation}
its value at $z$.

%This is a simple transformation, as it is possible to calculate one 
%power spectrum, and to calculate the mass variance at redshift of zero 
%before merely multiplying by the factor $d(z)$ to procure the mass 
%variance at another epoch. Figure \ref{fig:massfuncz} shows the effect 
%of changing redshift on the HMF using a simple Sheth-Tormen fit.

%\begin{figure}
%  \label{fig:massfuncz}
%  \centering
%  \includegraphics[width=\linewidth]{hmfz.pdf}
%  \caption{The halo HMF calculated using the Sheth-Tormen approach, 
%    at the present day in a fiducial CDM model, and at redshifts of $z=1$ and 
%    $z=5$.}
%\end{figure}

\subsection{Warm Dark Matter Models}

\begin{figure*}
  \label{densities}
  \centerline{
    \includegraphics[width=0.5\linewidth]{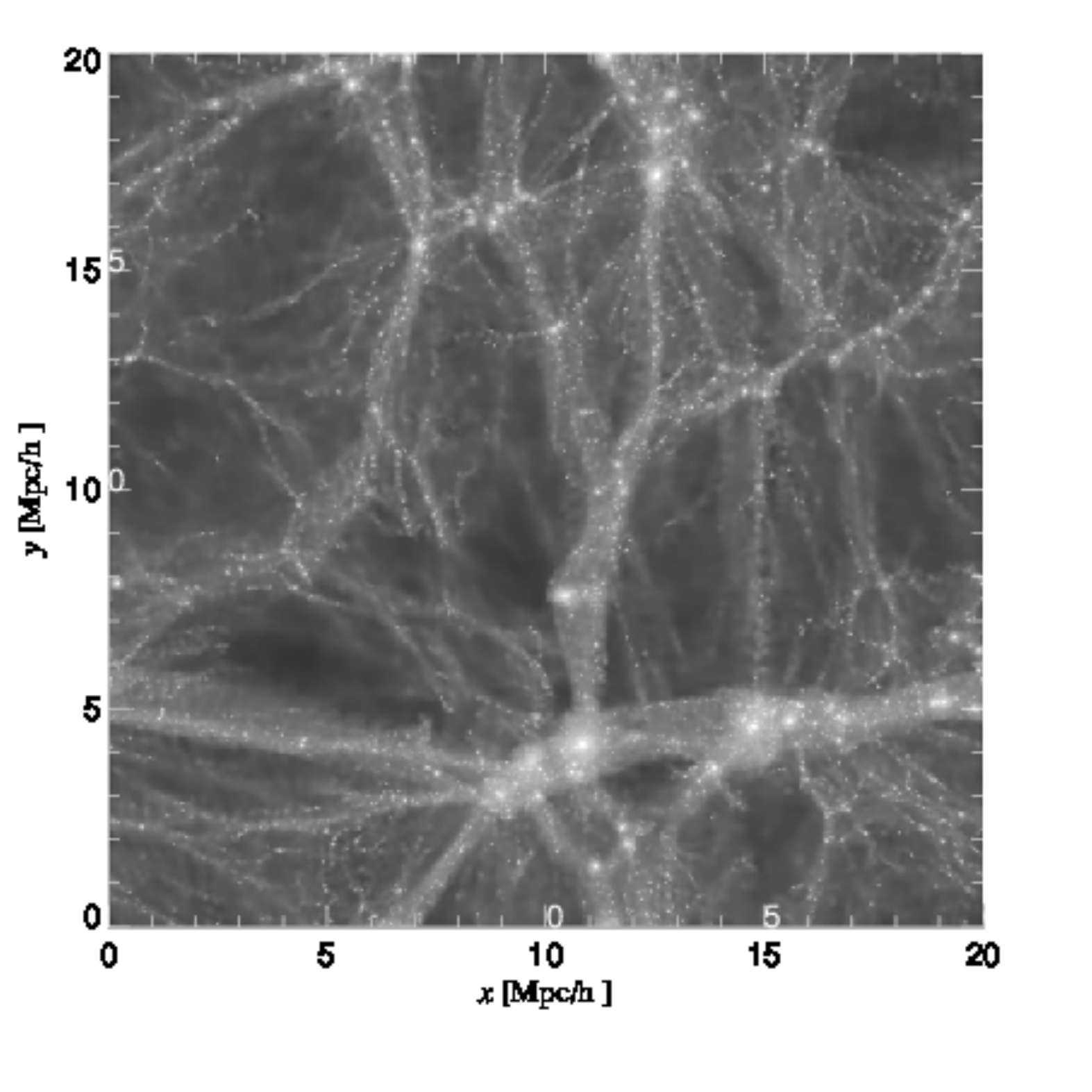}
    \includegraphics[width=0.5\linewidth]{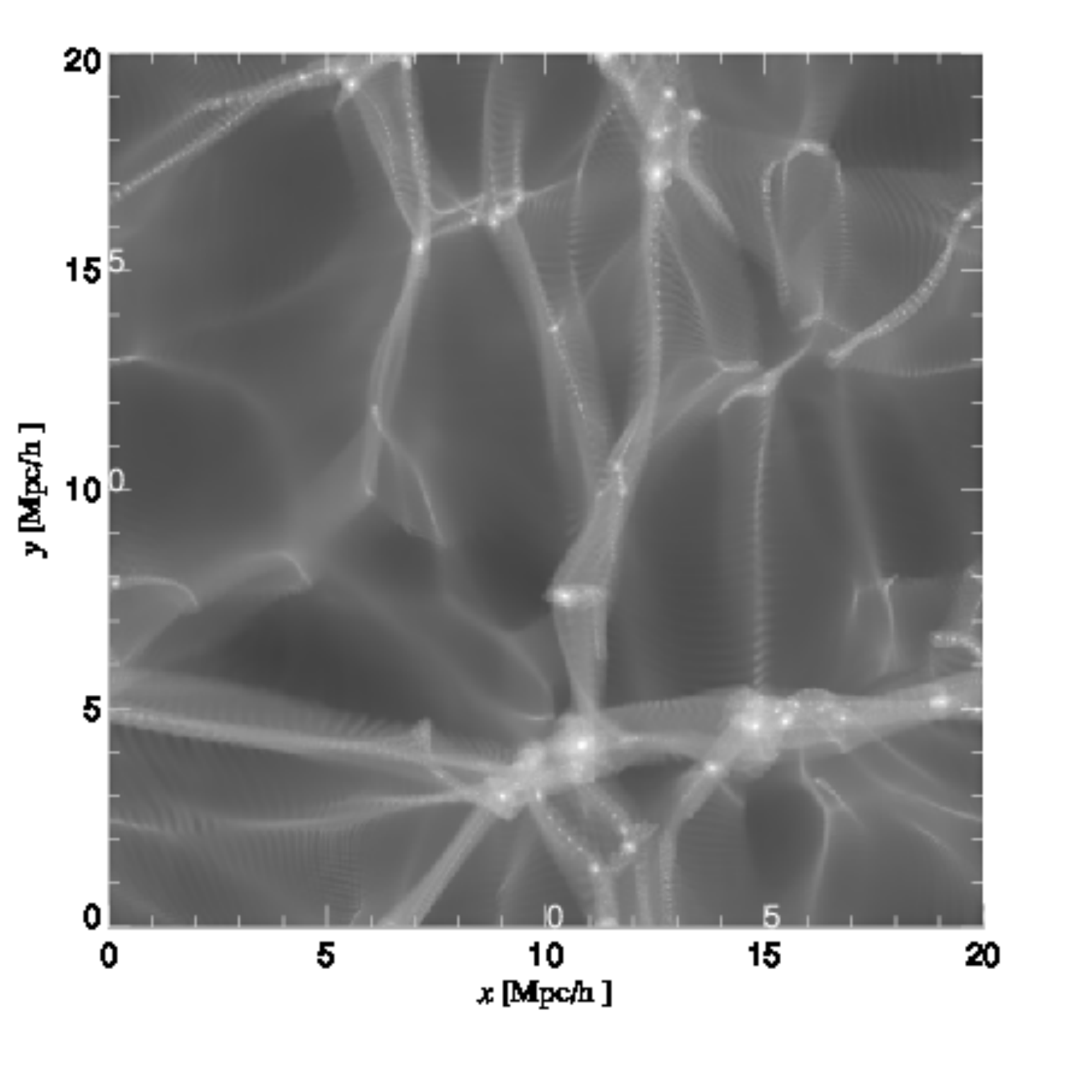}
  }
  \caption{Visual impression of the projected dark matter density in 
    a cosmological $N$-body simulations of a 20 $h^{-1} \rm Mpc$ box,
    modelling the growth of structure in a fiducial CDM model (left panel)
    and its WDM counterpart (right panel). For the WDM model we assume 
    a particle mass of $m_X$=0.5 keV/$c^2$. Note the absence of small-scale
    structure (i.e. low mass dark matter haloes) in the WDM run compared to the
    CDM run.}
\end{figure*}

The default in \verb|HMFcalc| is to calculate the HMF assuming an 
underlying Cold Dark Matter (CDM) model, but we have included the 
option to calculate the HMF in Warm Dark Matter (WDM) alternatives.
Small-scale power is suppressed in a WDM model compared to its CDM
counterpart, reflecting the free-streaming of WDM particles in the
early Universe, dispersing small-scale density perturbations that
would otherwise collapse via gravitational instability to form 
low-mass dark matter haloes. The influence of the dark matter model
on structure formation is readily apparent in Fig~\ref{densities},
which shows the projected dark matter density at $z$=0 in a 
20 $h^{-1}\rm Mpc$ patch in a fiducial CDM model (left panel) and 
in a WDM model assuming a WDM particle mass of $m_{X}=0.5 {\rm keV}/c^2$.
Although the large-scale features (filaments and voids, positions of
the most massive haloes) are the same in both runs, a wealth of 
small-scale structure -- low mass dark matter haloes -- is evident in 
the CDM run but not in the WDM run.

The standard approach to capturing the effect of WDM on the power
spectrum is to filter the corresponding CDM power spectrum on small 
scales by an additional transfer function, i.e. $P(k) = P_{CDM} {T_k^{X}}^2$.
The transfer function of \citet{Bode2001} is widely used, and it is the
one we adopt in \verb|HMFcalc|; here
\begin{equation}
	\label{eq:wdmtransfer}
	T_k^X = \left(1+(\alpha k) ^{2\nu}\right)^{-5/\nu},
\end{equation}
with $\nu =1.2$ and
\begin{equation}
\alpha = 0.048\left(\frac{\Omega_X}{0.4}\right)^{.15}\left(\frac{h}{.65}\right)^{1/3}\left(\frac{1}{m_X}\right)^{1.15}\left(\frac{1.5}{g_X}\right)^{.29}
\end{equation}
with $\Omega_X$ the current fractional density of the WDM particle (this 
can be taken as equivalent to the CDM density $\Omega_{cdm}$ in a 
single-species WDM model), $m_X$ is the particle mass in keV, and $g_X$ 
controls the abundance of the species relative to photons and has the 
fiducial value of 1.5 for a light neutrino. 

\begin{figure}
  \label{fig:massfuncwdm}
  \centering
  \includegraphics[width=\linewidth]{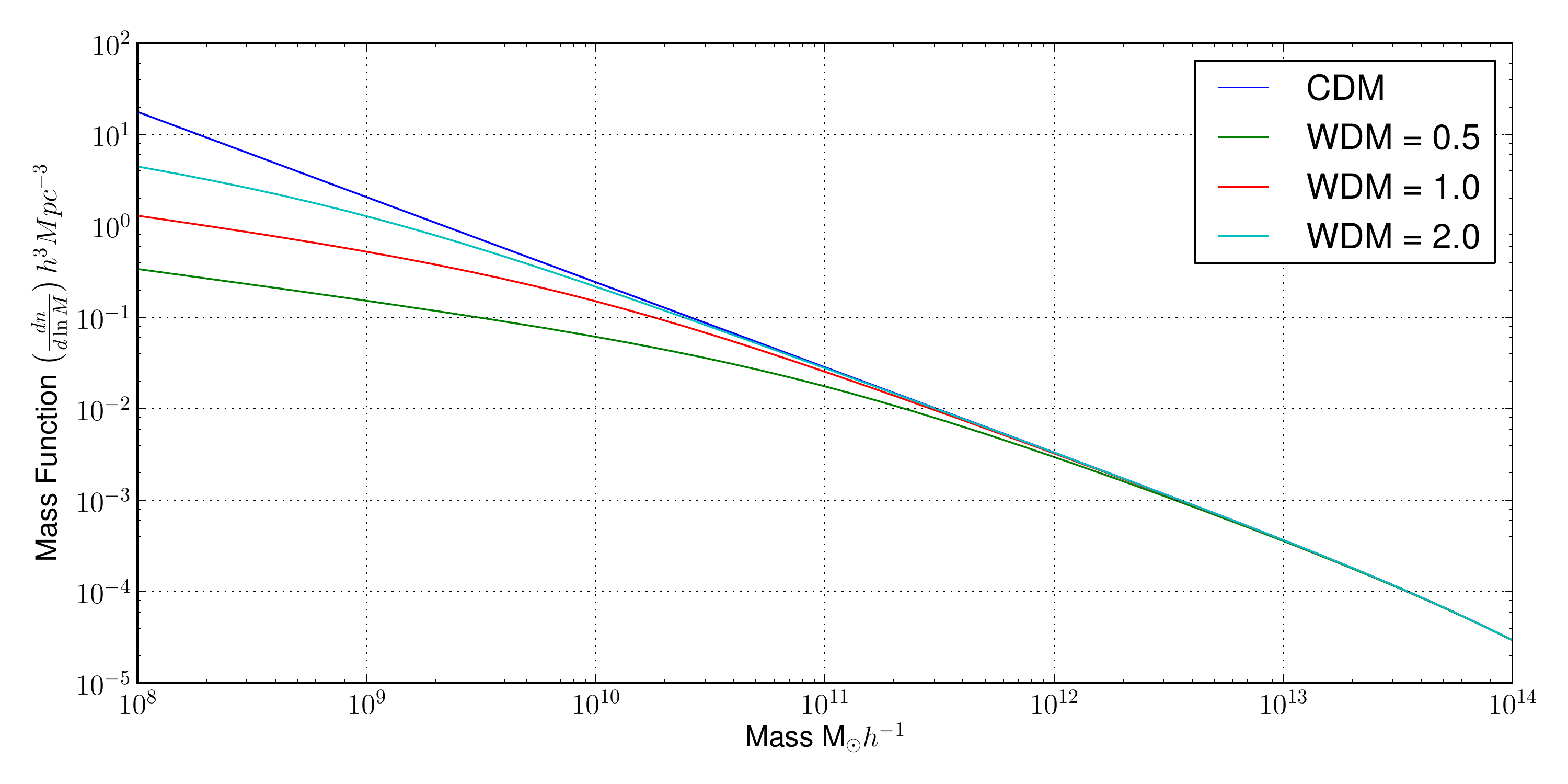}
  \caption{The HMF calculated assuming a fiducual CDM model and 
    three different WDM models with particle masses of $m_X$=0.5, 1 and
    2 keV/$c^2$. Note the suppression of the number density of lower mass
    haloes.}
\end{figure}

By default in \verb|HMFcalc|, we assume that $\nu$ and $g_X$ are set to 
their fiducial values and allow only a single-species model; the only 
free parameter that we allow is $m_X$. Fig \ref{fig:massfuncwdm} shows 
a fiducial CDM and WDM HMFs for $m_X$= 0.5, 1 and 2 keV/$c^2$ -- as the 
WDM particle mass decreases, so too does the number density of lower mass 
haloes, as we would expect. However, as we note in \S~\ref{sec:future}, 
there have been recent developments in the modelling of Warm Dark
Matter models \citep[e.g.][]{Schneider2013,Pacucci2013,Benson2013} that we
will incorporate in future versions of \verb|HMFcalc|.

\section{Implementation}
\label{sec:implementation}

\verb|HMFcalc| has been designed to be visually intuitive, easy to learn and 
use, flexible and extensibility. These requirements guided our implementation;

\begin{itemize}
  
\item The code is written in Python, an extremely versatile language 
  that is widely used in the astronomical community.
  
\item The user interface is a web-app frontend, which is designed to work
  across multiple platforms and for its ease of use.

\item The web-framework is provided by Django. It is flexible, easy to use 
  and being written in Python, it integrates easily with the 
  \verb|hmf| backend.
  
\item The whole code has been made open-source at 
  \url{https://github.com/steven-murray/HMF}. This way the code 
  can be examined and updated, and also understood by those 
  wishing to learn about the HMF.
  
\end{itemize}

We have also chosen to take an object-oriented approach, which makes sense because 
many of the HMF calculations share attributes. Class structures\footnote{Objects 
  to which attributes and methods may be attached.} provide a way to define common 
attributes and make them available throughout all methods of the class, 
which in our case enables faster updating of parameters. 

In the following subsections we describe in detail the backend \verb|hmf| 
(\S~\ref{sec:calcs}), before introducing the web-application (hereafter web-app) 
frontend \verb|HMFcalc|.

\subsection{\texttt{HMFcalc}'s engine, \texttt{hmf}}
\label{sec:calcs}

The calculation of the HMF is performed using \verb|hmf|, which
consists of three python modules; 
\begin{enumerate}

\item \verb|tools.py| contains functions necessary for e.g. importing or 
  generating (as required) the transfer function, calculation of the mass 
  variance, interpolating the transfer function etc...;

\item  \verb|cosmography.py| contains functions that calculate cosmological 
  distances and other cosmographic quantities;

\item \verb|hmf.py| contains a single class dedicated to 
  computing the HMF, as well as quantities such as the power spectrum, the 
  mass variance, etc...

\end{enumerate}
The handling of the transfer function in \verb|tools.py| is done in one of two ways.
If a filename is provided, the relevant file will be read and used as the transfer 
function. This file must be either produced by \verb|CAMB| or in \verb|CAMB| format. 
If no filename is provided, \verb|CAMB| is called via 
\verb|pycamb|\footnote{\url{https://github.com/joezuntz/pycamb}}, an 
open-source python wrapper for \verb|CAMB|. This enables efficient 
shared-memory multi-core calculation of the transfer function for a range 
of input parameters. The \verb|tools.py| module also contains a 
function that checks the bounds on $kR$, as discussed in \S\ref{sec:theory}.

\verb|hmf.py| is the central engine of \verb|hmf|, containing a 
single class named \verb|Perturbations| that contains all attributes needed
to calculate the HMF (the power spectrum, limits in $k$, the redshift, 
cosmology and mass variance etc.) and all the methods needed to calculate 
these quantities. 

The constructor for the \verb|Perturbations| class
initialises an `empty' object with a range of cosmological and 
numerical parameters set. The various quantities associated with the object
are calculated only when necessary, by using the \verb|@property| decorator 
that allows the getting, setting and deleting of class properties in a flexible 
manner. These include the power spectrum (\verb|power|), mass variance (\verb|sigma|), 
growth factor (\verb|growth|), effective spectral index (\verb|n_eff|), fitting 
function (\verb|fsigma|) and most importantly the differential and cumulative mass 
functions (\verb|dndlnm| and \verb|ngtm|). 
 
Each fitting function is defined as a single method within the 
\verb|Perturbations| class. However, the \verb|dndlnm| method 
chooses between them using a dictionary indexed by string identifiers 
(such as `ST' for \citet{Sheth2001} or `Reed03' for \citet{Reed2003}). 
Extending the range of fitting functions is as straightforward as writing a method 
containing the function definition, and registering it in the dictionary 
with a unique string identifier. However, users also have an option to pass a 
function in the format of a simple string that the class will parse and use as the fitting function.

Algorithm \ref{al:simplecalc} summarises the steps necessary for the calculation
of the differential HMF in \verb|hmf|. Interpolation of the transfer function 
is performed with linear splines on the function in log-space, in which the function is 
almost perfectly linear over much of its range. This is consequently re-gridded 
between the given $k$-bounds in 4097 steps, to ensure sufficient accuracy for
for romberg integration. Calculation of the cumulative HMF, $n(>M)$, is 
likewise performed using linear spline interpolation and extrapolation up to 
$M = 10^{18}{\rm M}_{\Sun}h^{-1}$ and consequent romberg integration using 
4097 steps defined for each mass.
  
Users may want to re-calculate the HMF for different cosmological parameters,
redshifts or fitting functions.
To optimise this process, \verb|hmf| tracks dependencies of quantities through chaining
immediate links. For example, algorithm \ref{al:simplecalc} demonstrates that the 
unnormalised power spectrum at the current epoch is dependent on the transfer function
(and therefore all cosmological and numerical parameters involved in that calculation), 
and also the spectral index, but it does not depend on redshift, normalisation, or overdensity
(either virial or critical). Any variable can be updated by calling the \verb|update()| 
method with relevant keyword arguments; quantities that depend \textit{directly} on the updated variables are deleted, 
which triggers deletion of quantities that depend directly on these and so forth. The next time any of these quantities are required, they are automatically 
re-calculated, using the updated variable. This system ensures that only the 
quantities that need to be re-calculated are re-calculated.

\begin{algorithm}
  \caption{Calculate Halo HMF}
  \label{al:simplecalc}
  
  \begin{algorithmic}
    \If {No Transfer File}
    \State $k$,$T$ $\gets$ \verb|CAMB|()
    \Else
    \State $k, T \gets$ ImportTransferFile()
    \EndIf			
    
    \State $T_f() \gets$ Interpolate $T$
    
    \If {Extrapolating Power Spectrum}
    \State Set New $k$ Bounds				
    \EndIf
    
    \State Re-grid  $k$ between bounds 
    \State $T \gets T_f(k)$			
    
    \State $	P_0(k) = k^n  T^2$ \Comment Eq~\ref{eq:powerfromtransfer}
    
    \State $\hat{\sigma_8} \gets \sigma(M=8,z=0)$  \Comment {Eq~\ref{eq:massvariance}} 
    \State $P(k) \gets   \left(\frac{\sigma_8}{\hat{\sigma_8}}\right)^2P_0(k) $ \Comment {Normalise}
    
    \If {WDM}
    \State Find $T_{WDM}(k)$ \Comment{Eq~\ref{eq:wdmtransfer}}
    \State $P(k) \gets P(k) * T_{\mathrm{WDM}}^2$ \Comment{Eq~\ref{eq:wdmtransfer}}
    \EndIf 
    
    \State $\sigma_0(M) \gets \sigma(M,z=0)$   \Comment{Eq~\ref{eq:massvariance}}
    \State $K \gets \frac{d \ln(\sigma)}{d \ln(M)}$   \Comment{Eq~\ref{eq:dlndlnm}}
    
    \State Find $d(z)$ \Comment{Eqs~\ref{eq:dofz} - \ref{eq:HofA}}
    
    \State $P(k) \gets P_0(k) d^2(z)$ \Comment {Apply Linear Growth}					
    \State $\sigma(M,z) \gets \sigma_0(M)d(z)$ \Comment{Apply Linear Growth}		
    \State Calculate $f(\sigma)$ with appropriate form
    \State HMF $\gets \frac{\rho_0}{M}*f(\sigma)*K$ \Comment {Eq~\ref{eq:hmf}}
    
  \end{algorithmic}
\end{algorithm}

This updating method allows for the optimal calculation of many 
HMFs for varying parameters in an efficient and error-free 
manner. We have used it in \citet{Murray2013}, generating 20,000 varying 
HMF realisations for each of 11 different fitting functions. On average
the cascading method reduces processing time by 50\% in tests where each 
available parameter is changed one time, although the size of reduction
depends on which parameters are most often modified (e.g. changing only
redshift reduces computations on average by more than 94\%).

\subsection{\texttt{HMFcalc}'s web interface and the Django Framework}

\begin{figure*}
  
  \centering
  \includegraphics[width=\linewidth, trim=1cm 8.2cm 0.25cm 8.0cm, clip=True]{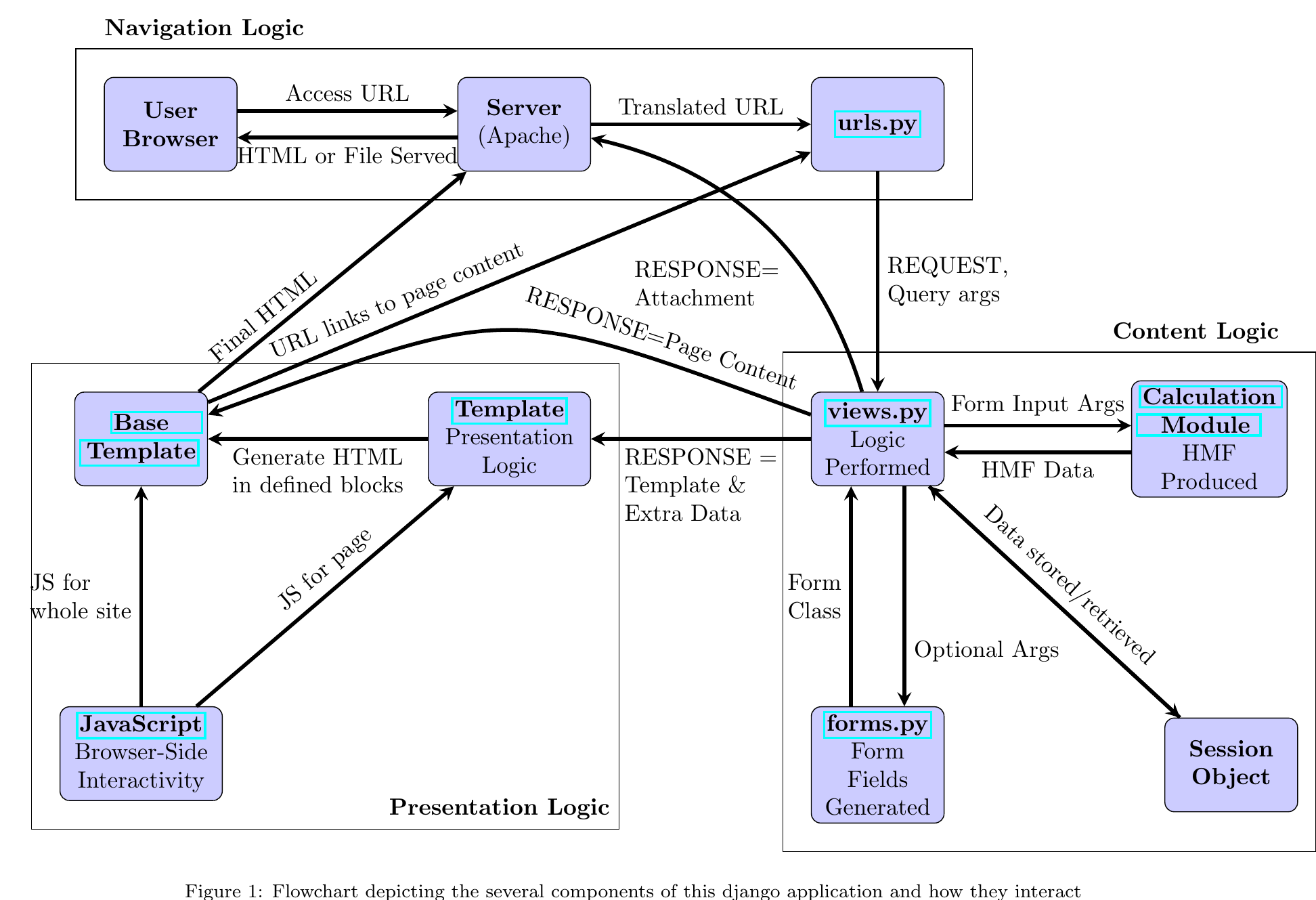}
   \label{fig:flowchart}
\end{figure*}

Django provides a web-framework tuned to fast development and 
performance. Although developed with online newsrooms in mind, Django 
is easily adapted to scientific applications. A scientific Django web-app 
will generally be built around four parts -- urls, views, templates and 
calculation -- and Fig \ref{fig:flowchart} depicts the flow of information 
within this framework\footnote{Note that the `Calculation Module' node is 
  a black-box for the actual calculations using the \texttt{hmf} backend, 
  which has been described already.}. The user initiates the information 
flow by pointing their browser at a particular URL. This URL is transformed 
to an address and contacts the Apache server on which \verb|HMCcalc| is 
operating. The server translates the URL for input to the app's own 
\verb|urls.py| module, which parses it and uses the information to choose a \verb|view| 
function from the \verb|views.py| module. This function may also be 
passed extra arguments from the input URL, based on query strings. The 
\verb|view| chosen does all the logic, and dictates \textit{what} should 
be seen on the final page. In the case of \verb|HMFcalc|, one of the 
views accesses a \verb|form| within the \verb|forms.py| module for 
display. This form is passed through to a \textit{template}, which is an 
enhanced html file. The template controls the presentation logic, 
putting the form fields down in appropriate positions on the page (with 
the help of CSS). The final template is then passed through to the 
server to be displayed by the browser. 

We utilise a standard approach to the Django implementation. 
Nevertheless, we note some aspects of the design. Most of the views are 
defined as \textit{generic class-based views}. This enables simpler 
construction and more flexibility in function. It also supports the 
creation of dynamic tabs for the user's navigation, via the plugin 
\textit{django-tabination}. This implementation of the views is most 
important in the view that displays the main input form.

The view that displays the form is non-standard because there are two
variants of it -- one is for the initial creation of data, the other is 
to add more data. The forms for these differ in that the first lets the 
user choose a mass range (and grid), while the second does not -- it 
uses the mass range specified previously. To create this 
functionality in the simplest way, we define a base form view, which 
implements the same logic for each form (i.e. what to do with the data 
after it has been successfully entered). We then create two subclasses, 
one for creation and one for addition of data. Each of these passes 
different arguments to the form class, which dynamically adds or removes 
the mass range inputs. This construction also allows for each of these 
subclasses to define their own tab for user navigation.
 
Upon successful submission of the form, the base form view performs the 
main analysis. To do this, it takes the parsed input arguments and 
passes them to an external function that wraps around the
\verb|hmf| backend. For each combination of the input parameters, an HMF
is produced and saved as an entry in a python dictionary with a key
unique to the combination of parameters. This process is greatly simplified by
using the \verb|update()| method of the \verb|hmf| package.

To enable later retrieval of the data for plotting and/or addition 
of extra data, the input parameters and calculated HMFs are 
saved to the \verb|session| object, which runs continuously throughout the 
course of a user's session. It can be accessed by any view and 
provides a convenient way to transfer data. This is used as sparingly as 
possible, however, because it increases overall memory usage.

In terms of presentation, we utilise Django's templating system, combined with
the \textit{bootstrap}\footnote{\url{http://twitter.github.io/bootstrap/}} CSS framework. This system allows for \textit{template inheritance} -- base templates 
can be written that define certain blocks that sub-templates may alter. 
This is convenient because we define a single base template for the 
whole site, which provides the general layout, including the navigation 
bar. The template for each page needs only to specify the unique 
content, which simplifies the design greatly. We use the \textit{bootstrap} framework
as it provides a simple means to create an aesthetically pleasing design.

Output and presentation of the calculations is done by means of the 
creation of figures and generation of downloadable ascii data.
For both of these options, all files and resources are served in memory --
nothing is written to disk on the server. This enhances 
performance and increases simplicity server-side. We have chosen to use 
\verb|matplotlib| \citep{Hunter2007}, the standard scientific plotting library of Python, 
to drive the plotting for the application. This is used to produce PNG
images for dynamic display on the web, and PDF images for optional 
download. Currently the plotting process is quite static -- the data 
itself is the only dynamic aspect. This may result in plots that are not 
drawn in the best possible way for a given set of parameters. However, 
it is expected that those who wish to plot many different combinations 
will likely download the data file and come to their own plotting 
solution.

\subsection{Usage}

Using \verb|HMFcalc| is straightforward -- generation of a HMF is as 
easy as pressing ``Calculate!" because the web form is populated with
default parameters. Submission of the form generates a page with a plot 
of the calculated HMF with relevant legend, as well as a drop-down menu 
for choosing the particular additional functions to be displayed. These 
include the mass variance, the fitting functions, the cumulative HMF and the 
power spectrum. Furthermore, an additional drop-down menu allows the user
to download the currently viewed plot, all calculated plots, or data 
files from the calculation. Below the plot additional cosmographic 
quantities that have been calculated are presented (e.g. age of Universe
at redshift $z$, distances, growth factors, etc...). In the following 
subsections we provide a glossary of parameters used in \verb|HMFcalc|.

\subsubsection{Parameters}

The first category of parameters are the `run parameters', which control the 
calculation (cf. Table \ref{tab:runparams}). Where multiple parameters are possible,
all combinations of parameters will be used in the calculation. This means that 
the number of HMFs generated can rise rapidly and so we place a limit on the total
number of possible combinations.

\begin{table*}
  \caption[]{Description of Run Parameters} 
  \label{tab:runparams}	
  \begin{tabu} to \linewidth{X[-2,l]X[-1c]X[8l]} 
    
    \toprule[0.04cm]
    \textsc{Parameter} & \textsc{Mult?}&\textsc{Description}\\ \midrule[0.03cm] %\endfirsthead
    %	\caption[]{(continued) } \\
    %	\midrule[0.03cm]
    %	\textsc{Parameter} & \textsc{Mult?}&\textsc{Description}\\ \midrule \endhead
    %		
    Redshift & \tick &  Epoch at which the HMF is calculated; default is $z$=0. 
    Affects HMF through growth parameter (cf. \S\ref{sec:evolution}), quick to 
    update. Some recent fitting functions contain explicit redshift dependence.\\ \midrule
    Virial Overdensity $\Delta_{\rm vir}$ & \tick & Defines the extent of a halo; default 
    is 178. Assumes haloes correspond to spherical overdensities of 
    $\Delta_{\rm vir}$ times the critical density at that epoch. Note 
    \citet{Tinker2008} and \citet{Watson2012} HMFs depend explicitly on 
    $\Delta_{\rm vir}$.\\ \midrule
    WDM Mass $m_X$ & \tick & WDM particle mass in single-species models, in units of 
    keV/$c^2$; default is 0. This is used to calculate the WDM transfer function and
    consequently the power spectrum (cf. Eq~\ref{eq:wdmtransfer}). \\ \midrule
    Fitting Function & \tick & Defines fraction of mass collapsed at a given mass 
    scale. See \S\ref{sec:refinedfittingfunctions} for more details.  \\* \midrule
    Custom Fitting Function & \cross & User-defined fitting function, written in Python
    syntax as a string, with independent variable named $x$ corresponding to the 
    mass variance; default is empty. To add more complex functions, the user will need 
    to edit the source code itself.\\ \midrule
    Extrapolate? & \cross & Enables/disables extrapolation the transfer function in 
    wavenumber $k$ beyond tabulated bounds; default is enabled. Assumes transfer 
    function is linear (in log space) near bounds and so can be easily extrapolated. \\ \midrule
    Min \& Max $k$ & \tick & Minimum and maximum wavenumber ($k$) for the power
    spectrum; defaults are $\sim10^{-8}$ and 2000 in units of $h \rm Mpc^{-1}$
    for accurate prediction (cf. \S\ref{sec:theory})\\ \midrule
    Min \& Max Mass & \cross & Defines minimum and maximum halo mass bounds for
    plotted HMF; defaults are $8$ and $15$ in units of 
    $\log_{10} h^{-1} \rm M_{\Sun}$.\\ \midrule
    Mass Bin Width & \cross & Defines logarithmic mass bin width; default is 0.05 dex.\\
    \bottomrule[0.04cm]
  \end{tabu}
\end{table*}

The second category of parameters are the `cosmological parameters' 
(cf. Table \ref{tab:cosmoparams}). 
The key parameter is the transfer function, which is invoked in one of 
three modes;
\begin{enumerate}
\item Pre-calculated (default); there are several transfer functions 
  pre-calculated for commonly used cosmologies, each produced at high 
  resolution by \verb|CAMB|.
\item Custom; the user may upload their own transfer function, assumed to
  have been produced either using \verb|CAMB| or in \verb|CAMB| format.
\item On-the-fly; \verb|CAMB| is run at invocation to produce the required
  transfer function.
\end{enumerate}
If a pre-calculated transfer function is chosen, all cosmological 
parameters are set as read-only on the form; if a custom transfer
function is used, the user must ensure that the cosmological parameters 
entered in the web-app match those used to generate the transfer function.

In Table \ref{tab:cosmoparams}, we use T, H and B to indicate whether or not
a parameter affects the transfer function (T), the HMF (H) or both (B). Note
that those that affect both need special care if the user uploads their 
own transfer function.

\begin{table*}
  \caption[]{Description of Cosmological Parameters}
  \label{tab:cosmoparams} 
  \begin{tabu} to \linewidth{X[-2,l]X[-1c]X[-6.5l]} 
    
    \toprule[0.04cm]
    \textsc{Parameter} & \textsc{Affects}&\textsc{Description}\\ \midrule[0.03cm] %\endfirsthead
    %	\caption[]{(continued) } \\
    %	\midrule[0.03cm]
    %	\textsc{Parameter} & \textsc{Affects}&\textsc{Description}\\ \midrule \endhead
    
    Transfer Function & -- &  Choose from drop-down menu. Either pre-calculated and 
    based on popular cosmologies or customised based on the available parameters.\\ \midrule
    HMF Labels & -- & Must be unique -- one for each parameter set plotted. Appears in
    plots as legends and in downloadable data as column headers.\\ \midrule
    $\delta_c$ & H & Critical overdensity for collapse. Although
    cosmology-dependent, we assume the Einstein-de Sitter value of 1.686 as default.\\ \midrule
    $n_s$ & H &  Spectral index of power spectrum -- see 
    \S\ref{sec:theory}). \\ \midrule
    $\sigma_8$ & H & Mass variance on scale of 8 $h^{-1} \rm Mpc$ at $z$=0 used to 
    normalise power spectrum -- (see \S\ref{sec:theory}).\\ \midrule
    $H_0$ & T & Hubble Parameter at $z$=0.\\ \midrule
    $\Omega_b$ & B & Baryon density at $z$=0.\\ \midrule
    $\Omega_c$ & B & Dark matter density at $z$=0.\\ \midrule
    $\Omega_{\Lambda}$ & B & Dark energy density at $z$=0.\\ \midrule
    \bottomrule[0.04cm]
  \end{tabu}
\end{table*}

\subsubsection{An Important Caveat}
\verb|HMFcalc| will calculate a HMF given any set of cosmological parameters
and fitting function. However, most HMFs are empirical fits to the results of
cosmological $N$-body simulations and are tied to specific cosmologies 
\citep[but see, e.g.][]{Jenkins2001,Courtin2010}); whether or not the HMF
is universal remains a topic of active research (see discussion in 
\citealt{Tinker2008}, \citealt{Bhattacharya2011} and \citealt{Murray2013}). For 
this reason we caution the careful user to compare the results of \verb|HMFcalc|
against N-body simulations of the appropriate cosmology.

\subsection{Comparison To Other Codes}
A small number of public codes are available for calculating the HMF -- eg.
\verb|genmf|\footnote{\url{http://icc.dur.ac.uk/Research/PublicDownloads/genmf_v1.01.tar}}, \verb|MF_Code|\footnote{\url{http://cosmo.nyu.edu/~tinker/massfunction/MF_code.tar}}, and the HMF calculator in the \verb|yt| project\footnote{\url{http://yt-project.org/}}.
We have tested \verb|hmf| against these codes and find excellent consistency.
However, we note that \verb|HMFcalc| has a number of advantages over these codes; it
is inherently visually interactive and cross-platform, it allows for dynamic accurate 
calculation of the transfer function with \verb|CAMB| and efficient and self-consistent 
parameter updates, and it can be easily extended by the user with bespoke fitting 
functions.
    
\section{Example Applications}
\label{sec:application}

We have designed \verb|HMFcalc| to be a useful tool for observers and 
theorists alike, and we expect the primary use of \verb|HMFcalc| to be the 
quick and straightforward generation of HMFs for different 
fitting functions and as a function of cosmological parameters and cosmic 
epoch. However, \verb|HMFcalc| should be a particularly useful aid to those 
who work with cosmological simulations, and we demonstrate how with 
two examples.

\subsection{Box Size for One Halo of Mass M}
\label{sec:howmany}

A cosmological simulation follows the growth of structure in a periodic 
cubic box of volume $L^3 {\rm Mpc}^{3}h^{-3}$ containing $N_p$ particles. In 
principle $L$ can be arbitrarily large, but normally it's set by 
requiring that the smallest resolved haloes contain a minimum number 
of particles, or equivalently the particle mass $m_p \propto L^3/N_p$. 
Given this restriction on $L$ enforced by $m_p$, it is interesting to 
ask how large the periodic box needs to be to contain at least one halo 
of mass $>M$. We require
\begin{equation}
  \label{eq:big_box}
  n(> M) L^3 \ge 1 \rightarrow L\ge n(> M)^{-1/3}.
\end{equation}
We have chosen to create a method in \verb|Perturbations()| (from 
the \verb|hmf| backend) to derive Eq~\ref{eq:big_box}. We add the 
following code to the class:

\lstset{language=Python,commentstyle=\color{Green4},keywordstyle=\color{Blue3},stringstyle=\color{Purple2},showstringspaces=false}
\begin{lstlisting}
@property
def how_big(self):
    try:
        return self.__how_big
    except:    
        return self.ngtm**(-1./3.)
    	
@how_big.deleter
def how_big(self):
    try:
        del self.__how_big
    except:
        pass
\end{lstlisting}
This is all that is needed to use \verb|hmf| to calculate 
Eq~\ref{eq:big_box}. Note the \verb|@property| decorator
ensures that the value of \verb|how_big| will change as the
value of \verb|ngtm| is modified. We extend the web interface
to output this information by adding the following to the function 
\verb|hmf_driver()| in \verb|utils.py| towards the end of the loops;

\begin{lstlisting}
if 'get_L' in extra_plots:
    mass_data["L(N=1)_"+getname(labels)] = 
                     pert.how_big
\end{lstlisting}
Here, the \verb|extra_plots| variable is a list that contains the names 
of the extra plots to be calculated (defined in the form later). 
The function is saved to the existing \verb|mass_data| array, under a 
label beginning with \verb|L(N=1)_|, which describes the ``box size 
needed for one halo". The \verb|Perturbations()| object has been 
previously initialized as \verb|pert|, and thus all we must do is access
its \verb|how_big| property. 

As we have noted, we must implement the choice 
of calculating this function on the form. To do this we add the 
entry

\begin{lstlisting}
('get_L', 'Box Size for One Halo'),
\end{lstlisting}
to the \verb|optional_plots| list variable in the \verb|HMFInput()| 
class in \verb|forms.py|. Its information will be passed to the form 
template, which will render it automatically in the `Optional Extra 
Plots' fieldset.

We also need to create a conditional entry for the function in the 
drop-down menu of the results page, so that we can view the plot. To do 
this we add the following code to the \verb|__init__()| method of the 
\verb|PlotChoices()| class of \verb|forms.py|:

\begin{lstlisting}
if 'get_L' in session_plots:
    extra_plots.append(('L',
         'Box Size for One Halo'))
\end{lstlisting}
Here we add the choice only if \verb|get_L| was checked on the initial 
input form. The first argument is the name of the choice internally, 
while the second is the label that the user sees.

The only remaining task is to create the plot definition for the 
function, in \verb|views.py|. We add the following code to 
produce the required axis labels and extract the $L(N=1)$ function from 
the whole dataset:

\begin{lstlisting}
elif plottype == 'L':
    keep=[string for string in mass_data \
          if string.startswith('L(N=1)_')]
    title="Box Size, L, for One Halo" 
    ylab= "Box Size, L (Mpc/h)"
    yscale = 'log'
\end{lstlisting}
and add the \verb|plottype| to the first list, \verb|mass_plots|, 
specifying that the $x$-axis will be based on the mass grid. Here the 
\verb|plottype| refers to the internal label from the 
\verb|PlotChoice()| form. The first line chooses all the columns 
whose labels begin with the specified string, corresponding to our 
custom function. Note that these code samples are merely indicative, nevertheless, the pattern will remain conceptually the same in future versions.

\begin{figure}
  
  \centering
    \includegraphics[width=\linewidth]{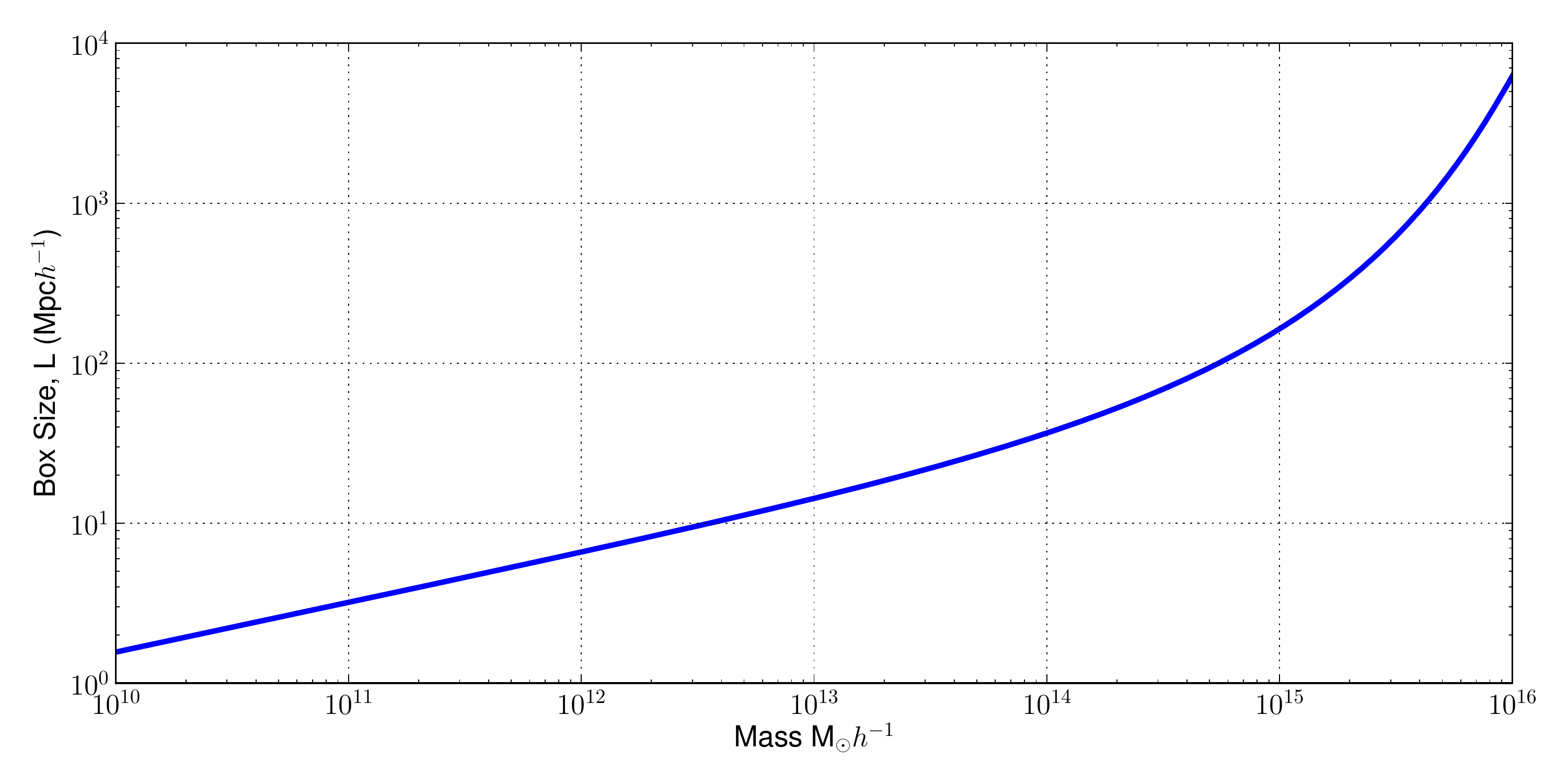}
    \caption{How large a box $L$ is required to form at least one halo of 
      mass $>M$? This shows the result for a WMAP7 cosmology at $z$=0.}
      \label{fig:howbig}
\end{figure}

\begin{figure}
  
  \centering
    \includegraphics[width=\linewidth]{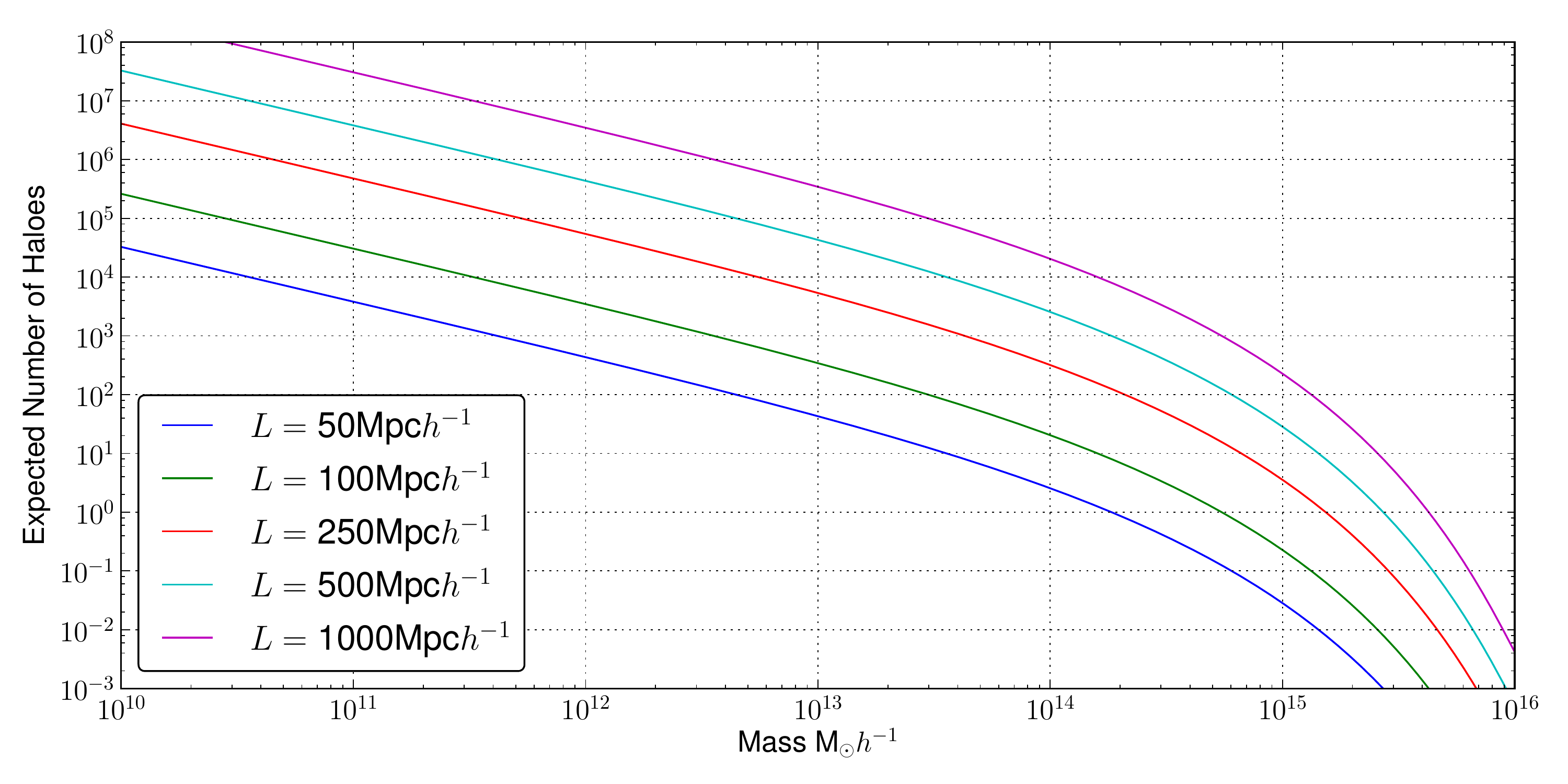}
    \caption{How many haloes do we expect of mass $>M$ in a simulation box of size $L$? This shows the result for a WMAP7 cosmology at $z$=0.}
    \label{fig:howmany}
\end{figure}

The result, for the WMAP7 cosmology at $z$=0, is shown in 
Fig~\ref{fig:howbig}; it is computed by default by the web-app.
We have also determined how many haloes are expected in
a simulation of box size $L$ (see Fig~\ref{fig:howmany}). This is
not currently implemented in the web-app, but it will be in future versions.

\subsection{The Impact of Finite Box Size}
\label{sec:finitesize}

Cosmological simulations model the growth of structure over cosmic
time by following the evolution of matter density perturbations in 
a periodic box. Because the box has a finite size $L$, the longest 
wavelength perturbation that can be reliably modeled will have a 
wavenumber $k_{\rm min}$ of 
\begin{equation}
  \label{eq:kmin}
  k_{\rm min} = \frac{2\pi}{L}.
\end{equation}
This means that the effect of perturbations with wavelengths longer 
than the box size cannot be captured, and implies that the mass 
variance we would expect (cf. Eq~\ref{eq:massvariance}) will be larger
than the one that we measure. This effect, and its influence on the HMF
at higher masses, has been noted in previous studies 
\citep[e.g][]{BaglaRay2005,PowerKnebe2006,Lukic2007}.

\begin{figure}
  \label{fig:finitebox}
  \begin{center}
    \includegraphics[width=\linewidth]{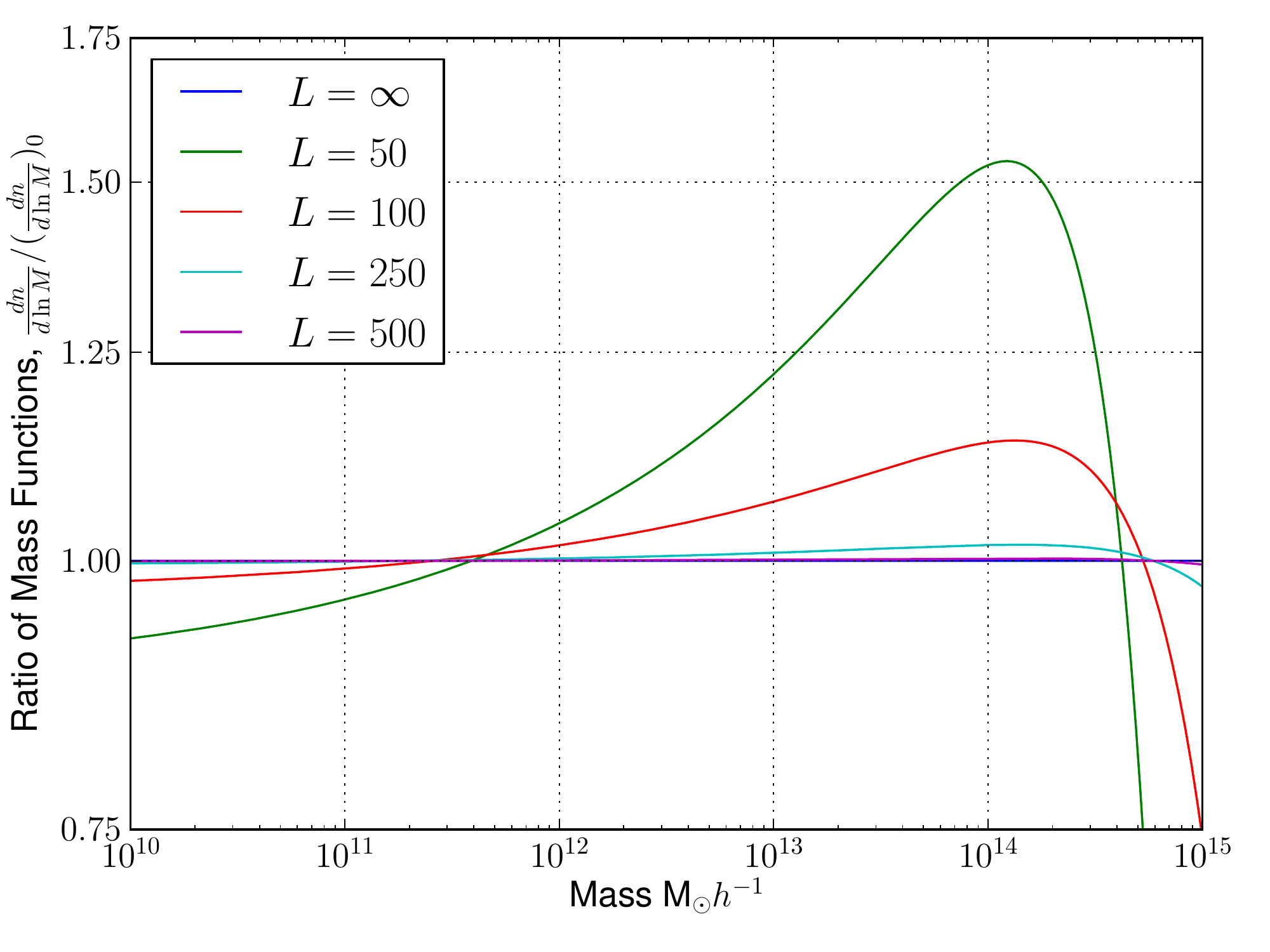}
    \caption{Artificial enhancement of the number density of haloes of mass
    $\sim 10^{14}{\rm M}_\Sun h^{-1}$ in smaller boxes.}
  \end{center}
\end{figure}

We can use \verb|HMFcalc| to quantify this effect for boxes of size 50, 
100, 250 and 500 $h^{-1} \rm Mpc$, corresponding to $k_{\rm min}$ = 
0.1257, 0.0628, 0.02513 and 0.01257 $h \rm Mpc^{-1}$ in Eq~\ref{eq:kmin}.
We use the inbuilt functionality of \verb|HMFcalc| to enter as input all 
of these minimum $k$ bounds at once, plus a baseline of $k_{\rm min} = 10^{-8}$ 
that corresponds to the limit of infinite box size. The result is shown in 
figure \ref{fig:finitebox}, which shows that the effect of box size is most
severe for the 50 $h^{-1}\rm Mpc$, with the number density of $10^14{\rm M}_\Sun h^{-1}$ mass 
haloes enhanced by $\sim$50\%; is weaker for the 100 $h^{-1} \rm Mpc$ box;
and is negligible for box sizes as large as 250 $h^{-1} \rm Mpc$. These results
are in good agreement with the findings of \citet{PowerKnebe2006}.
 
\section{Future Development}
\label{sec:future}

\subsection{Extending Range of Dark Matter and Dark Energy Models}
In the current version of \verb|HMFcalc|, we have considered only the 
fiducial Cold Dark Matter model and a straightforward extension to model 
Warm Dark Matter, using the framework set out in \citep{Bode2001}. In 
forthcoming versions of \verb|HMFcalc|, we shall update the framework to 
account for more recent developments in the modelling of Warm Dark
Matter models \citep[e.g.][]{Schneider2013,Pacucci2013,Benson2013}, as 
well as alternative dark energy models such as those explored in, for 
example, \citet{Francis2009a} and \citet{Courtin2010}.

\subsection{Calibrating Synthetic Galaxy Surveys}
``Halo Occupation Distribution'' (HOD) models parameterise how likely galaxies 
are to occupy dark matter haloes and provide a convenient framework 
for creating synthetic galaxy surveys. Formally, the HOD defines a 
probability $P(N|M)$ of finding $N$ luminous galaxies in a halo of 
mass $M$ \citep{Berlind2002}, and it allows a population of dark matter 
haloes to be mapped to a synthetic galaxy population whose abundance and 
spatial clustering can be tuned to reproduce the observed galaxy population. 
We will extend \verb|HMFcalc| so that we can take as input a given set of 
cosmological parameters, a dark matter and dark energy model, and a set of 
observational constraints (e.g. 2-point correlation function in different 
wavelength bands) and calibrate HOD parameters to reproduce the properties of 
the observed galaxy population. With this framework we will develop 
empirically-motivated synthetic galaxy surveys using both the outputs from 
both cosmological $N$-body simulations \citep[e.g.][]{Springel2006} as well 
as faster, more approximate schemes such as $2^{\rm nd}$ order Lagrangian 
perturbation theory \citep[e.g.][]{Manera2013}, vital in the coming era of 
large galaxy surveys.

\subsection{Dynamic and Adaptable User Interface}
We are developing \verb|HMFcalc| into a dynamic and adaptable online tool 
that will give the user finer and more varied control over the generation 
of data. For example, we would like users to be able to explore how 
variations in the cosmological parameters influence the HMF and to see 
the resulting HMF update in real-time. A similar framework will be used 
to allow users to explore how HOD-derived galaxy population observables vary
with changes in cosmological parameters and dark matter/dark energy parameters.
We would also like users to be able to explore how, for example, changing 
from the traditional top-hat window function (cf. Eq~\ref{eq:window}) to 
sharp-$k$ filter window functions \citep[cf.][]{Schneider2013} influence the 
predicted HMF, especially in WDM models. Such an online toolbox will provide 
an invaluable community resource for both observers and theorists alike.

\section{Summary}

We have presented \verb|HMFcalc|, a flexible, efficient and easy to use web 
application for calculating the dark matter halo mass function (HMF), a 
fundamental property of cosmological structure formation models. \verb|HMFcalc|
and its engine \verb|hmf| are implemented in Python, which offers a simple yet
powerful and extensible cross-platform method for building community software.
Observers and theorists alike should find \verb|HMFcalc| a valuable resource
to rapidly explore the effects of cosmological parameters, redshift and fitting 
function on the predicted HMF, while cosmological simulators will find it a useful 
tool for understanding, for example, the biases introduced by finite simulation 
volume. In future work we will extend \verb|HMFcalc| to give the user greater 
flexibility in how they interact with their data; we will broaden the range 
of dark matter and dark energy models considered; and finally, we will use the
framework provided by \verb|HMFcalc| to develop the infrastructure for fast
approximate methods for generating bespoke synthetic galaxy surveys.

\section*{Acknowledgments}
ASGR acknowledges support of a UWA postdoctoral research fellowship.
Part of this research was undertaken as part of the Survey Simulation 
Pipeline (SSimPL; {\small ssimpl-universe.tk}). The Centre for All-Sky 
Astrophysics (CAASTRO) is an Australian Research Council Centre of Excellence, 
funded by grant CE11E0090.

\bibliographystyle{elsarticle-harv}
\bibliography{library}

\end{document}